\documentstyle[12pt]{article}
\begin{document}
\begin{center} 
{\bf Scale Symmetry Breaking from the Dynamics of Maximal Rank 
Gauge Field Strengths}
\end{center} 

\bigskip 
\begin{center}
E. I. Guendelman\footnote{Electronic address: guendel@bgumail.bgu.ac.il}    
\end{center}  
\bigskip 
\begin{center}
{\it Physics Department, Ben Gurion University, Beer Sheva 84105, Israel}
\end{center} 
\bigskip

\begin {abstract}
Scale invariant theories which contain maximal rank gauge field strengths 
(of $D$ indices in $D$ dimensions) are studied. The integration of the 
equations of
motion of these gauge fields leads to the s.s.b. of scale invariance. The
cases in study are: i) the spontaneous generation of $r^{-1}$ potentials in
particle mechanics in a theory that contains only $r^{-2}$ potentials in
the scale invariant phase, ii) mass generation in scalar field theories,
iii) generation of non trivial dilaton potentials  in generally 
covariant theories,   
iv) spontaneous generation of confining behavior in gauge theories.
The possible origin of these models is discussed.   
\end {abstract}

\section{Introduction}
The idea that among the fundamental laws of physics, we have scale invariance 
(s.i.) as one of the fundamental principles appears as an attractive 
possibility. In its most naive realizations, such a symmetry is not a 
viable symmetry, however, since nature seems to have chosen some typical 
scales, so s.i. has to be broken somehow.

The origin of scales out of an originally scale invariant theory has a long
history. In most of the previous research, quantum effects that break s.i. 
have been investigated$^{1,2,3}$. The breaking of s.i. is crucial 
in the so called "induced gravity ideas". Here the Planck scale
can originate from the s.i. breaking, which can be explicit $^{4}$ 
(from an assumed scalar field potential which contains a mass parameter) or
may have its origin in quantum fluctuations  $^{5,6}$ .

While one may look at quantum effects for the origin of scale symmetry
breaking, there is also a way to achieve this at the classical level in
certain models (as we will discuss in the last section, the "classical"
scale symmetry breaking effect may be nevertheless related to some 
infrared singular behavior of an associated quantum theory in some cases).

In the first scale invariant model of this kind$^{7}$ , a metric, a dilaton
field and a "measure field" $\Phi$ are introduced. The unusual modified
measure$^{8}$ is an object that has the same transformation properties under
general coordinate transformations as  $\sqrt{-g}d^{4}x$ 
($g = det(g_{\mu \nu})$). $\Phi$ is a density built out of degrees of freedom
independent of the metric. For example,  given a four index field strength
(in four dimensions) 
$F_{\mu \nu \alpha \beta} = \partial _{[\mu} A _{\nu \alpha \beta ]}$,
the measure field $\Phi$ is defined as 
$\Phi = \epsilon ^{\mu \nu \alpha \beta} F_{\mu \nu \alpha \beta}$ .
One may consider the field strength to be  
composed of elementary scalars$^{9}$ . Four in the case of a four index
field strength, as it was done in Refs. 7, 8.

The scale symmetry breaking in Ref. 7 comes when considering the integration
of the equation of motion of the four index field strength, which introduces 
an arbitrary scale in the equations of motion.

The modified measure idea has been applied to dynamical generation of string  
and brane tensions$^{10}$ , to cosmological$^{11}$ and to the fermion family  
problem$^{12}$.

Such kind of "maximal rank" (four index in four space-time dimension)
Gauge field strength has a simple dynamics and were introduce in the context
of non scale invariant theories by several authors$^{13}$ .

Here we want to extend the use of "maximal rank" field strengths for the 
purpose of s.s.b. of scale invariance in models not involving gravity , 
including models in i) particle mechanics, where $\frac{1}{r}$ potentials 
appear after s.s.b. of s.i. given that only  $\frac{1}{r^{2}}$ potentials 
appear in the unbroken phase, ii) Field theory  models 
where non trivial scalar field potentials appear 
with s.s.b. of s.i. and may be also of internal symmetries, 
iii) the spontaneous generation of 
confining behavior in gauge theories. For the sake of completeness, we review
also the s.s.b. of s.i. in the generally covariant theory studied in Ref. 7
(section 4).

\section{A particle mechanics example} 

Let us start our discussion by considering what is possibly the simplest 
system displaying scale invariance, a non relativistic particle subjected to
an inverse square potential, i.e.
 
\begin{equation}
S = \int L dt,  L = \frac{1}{2} (\frac{dr}{dt})^{2} - \frac{K}{2r^{2}}
\end{equation}

where $K$ is a constant.

It is then straightforward to verify that the following scale transformation 
is a symmetry of (1),

\begin{equation}
r(t) \rightarrow r'(t) = \lambda ^{-\frac{1}{2}}  r(\lambda t)
\end{equation}

An equivalent form to (1) is obtained by considering the lagrange multiplier
$\omega$ in the new action

\begin{equation}
S = \int L dt,  
L = \frac{1}{2} (\frac{dr}{dt})^{2}  
+ \frac{1}{2} \omega ^{2} -  \omega \frac{K^{\frac{1}{2}}}{r}
\end{equation}  

then s.i. is obtained if in addition to (2) we also transform $\omega$ as in
 $\omega(t) \rightarrow \omega '(t)$
$ = \lambda ^{\frac{1}{2}} \omega(\lambda t)$
here  $\omega(t) $ is treated as a new dynamical variable. The equation of
motion obtained from the variation of $\omega(t)$ is

\begin{equation} 
\omega(t) = \frac{K^{\frac{1}{2}}}{r}
\end{equation}    

replacing back into (3) we obtain the action (1). Replacing $\omega(t)$ 
back into (3) is a legitimate operation because (3) is a constraint equation.
In any case, it is very much obvious that (1) and (3) lead to identical 
equations of motion.

We now consider a generalization of (1) or (3) which will give the 
solutions of (1) or (3) as a particular choice of the initial conditions.
This involves the consideration of the "maximum rank" field strength, which 
in one dimension means considering  $\omega(t) = \frac{d \sigma}{dt}$
and consider

\begin{equation} 
S = \int L dt,   
L = \frac{1}{2} (\frac{dr}{dt})^{2} + 
\frac{1}{2} (\frac{d \sigma}{dt})^{2} -                             
\frac{d \sigma}{dt} \frac{K^{\frac{1}{2}}}{r}
\end{equation}

s.i. is obtained if $r$ transforms as in (2) and $\sigma$ transforms as

\begin{equation}
\sigma(t) \rightarrow \sigma'(t) = \lambda ^{-\frac{1}{2}} \sigma (\lambda t)    
\end{equation}

Notice the existence of the additional global symmetry
\begin{equation}
\sigma(t) \rightarrow \sigma(t) + constant 
\end{equation}

Now the equation of motion that is obtained from the variation with respect 
to $\sigma$ is
\begin{equation}
\frac{d}{dt} (\frac{d \sigma}{dt} - \frac{K^{\frac{1}{2}}}{r} ) = 0
\end{equation} 

which can be integrated to give

\begin{equation}
\frac{d \sigma}{dt} = \frac{K^{\frac{1}{2}}}{r} + M    
\end{equation}

where $M$ is an integration constant. It corresponds to the conserved charge 
associated to the internal symmetry (7). As we will see the constant 
$M$ leads to s.s.b. of scale symmetry. One may notice the interesting
fact that the presence of a non trivial charge can lead to s.s.b $^{14}$ .
Equation (9) cannot be used in order to solve $\frac{d \sigma}{dt}$ and then
replace this value in eq. (5), that procedure leads to incorrect results.

The equation of motion associated with the variation with respect to $r$
(treating $\frac{d \sigma}{dt}$ in (5) as independent of r, as should be
done in the correct variational principle)
gives
 
\begin{equation}
\frac{d^{2}r}{dt^{2}} = 
- \frac{d \sigma}{dt} \frac{\partial}{\partial r} ( \frac{K^{\frac{1}{2}}}{r})
\end{equation} 
 
and using (9) we get,

\begin{equation} 
\frac{d^{2}r}{dt^{2}} = \frac{K}{r^{3}} + \frac{ MK^{\frac{1}{2}} }{r^{2}} 
= -\frac{\partial}{\partial r} V_{eff} 
\end{equation} 
  
where 
\begin{equation}
V_{eff} =  \frac{1}{2} \frac{ K }{r^{2}} + \frac{ MK^{\frac{1}{2}} }{r}
\end{equation}

We see that the constant of integration $M$ leads to a 
$\frac{1}{r} $ term in the effective potential.  
Notice that if we had added an explicit $\frac{1}{r} $ term in the action
(1), this would have explicitly violated scale invariance. Instead, we have 
now obtained the breaking of scale invariance after the integration of the
equations of motion. $M \neq 0 $ means indeed s.s.b. of scale invariance.

The spontaneous generation of a  $\frac{1}{r} $ term in the effective 
potential, in a theory where such a potential does not exist in its 
scale invariant phase, resembles very much the idea of induced 
gravity$^{4,5,6}$ , where the Newtonian gravity (the analogous of a 
$\frac{1}{r} $ term in the effective potential here) appears as a result 
of the breaking of s.i..

\section{Scalar Field Theory with "Maximal Rank" Gauge Fields}
We  consider first the scale invariant model, formulated in a
flat four dimensional space-time:

\begin{equation}
S = \int L d^4 x, 
 L = 
\frac{1}{2} \frac{\partial \phi}{\partial x^{\mu}}
\frac{\partial \phi}{\partial x_{\mu}}  - \frac{1}{2} \lambda \phi^4  
\end{equation}

Such a model is invariant under the scale transformation
\begin{equation}
\phi(x)  \rightarrow \phi'(x) = \lambda \phi(\lambda x)
\end{equation}  

As in the previous section, we can consider instead of (13) the following
equivalent action
\begin{equation}
S = \int L d^4 x,
 L = \frac{1}{2} \frac{\partial \phi}{\partial x^{\mu}}
\frac{\partial \phi}{\partial x_{\mu}}  + \frac{1}{2} \omega ^{2} 
- \omega \lambda ^{\frac{1}{2}} \phi ^{2}
\end{equation} 

The forms (15) and (13) are equivalent. Solving for $\omega$ from the 
equation of motion of $\omega$ and inserting back into (15) is a
valid manipulation. In any instance it is easy to check that (13) and
(15) are totally equivalent.

Consider now the case when we keep the form (15), but where now 
$\omega$ is not a fundamental field, but rather is given in terms of a
"Maximal Rank" Gauge Field, i.e.,
\begin{equation}
\omega = \epsilon ^{\mu \nu \alpha \beta}
\partial _{[\mu} A _{\nu \alpha \beta ]}     
\end{equation}       

$\partial _{[\mu} A _{\nu \alpha \beta ]}$ being the "Maximal Rank" 
Gauge Field for the four dimensional case.
We have now the invariance 
$A _{\nu \alpha \beta} \rightarrow$ 
$ A _{\nu \alpha \beta}  + \partial _{[\nu} \Lambda_{\alpha \beta]}$ .

The equation of motion associated with the variation of 
$A _{\nu \alpha \beta}$ is now
\begin{equation}
\epsilon ^{\mu \nu \alpha \beta}
\partial_{\beta} ( \omega - \lambda ^{\frac{1}{2}} \phi ^{2}) = 0 
\end{equation}

which is integrated to
\begin{equation}
\omega = \lambda ^{\frac{1}{2}} \phi ^{2} + M
\end{equation}

where $M$ is a space time constant that will be responsible for 
the s.s.b. of scale invariance.
Once again, as in the point particle case, we can see the effect of $M$
by studying the other eqs. of motion, in this case, that of $\phi$,
\begin{equation}
\partial_{\mu} \partial^{\mu} \phi = -2 \omega \lambda^{\frac{1}{2}} \phi 
\end{equation}

and by using (18), we get,
\begin{equation}
\partial_{\mu} \partial^{\mu} \phi = -2 \lambda \phi^{3} -
2M \lambda ^{\frac{1}{2}} \phi = -\frac{\partial V_{eff}}{\partial \phi}
\end{equation} 

where 
\begin{equation}   
 V_{eff} = \frac{1}{2} \lambda \phi ^{4} + 
M\lambda^{\frac{1}{2}} \phi^{2}
\end{equation}

We see that the constant of integration $M$, which is associated to 
s.s.b. of scale invariance is responsible for spontaneous generation of 
$mass^{2}$ which equals $2M\lambda^{\frac{1}{2}}$, notice that if 
$M<0$, we obtain also s.s.b. of the $\phi \rightarrow -\phi$ reflection 
symmetry, because $\phi $ gets an expectation value.

\section{Spontaneous Breaking of Scale Symmetry in a Generally Covariant
Model}
This section gives an example of work on scale invariant generally covariant
theories. There is no attempt to go through all the work done in this
area, but rather present one example where the "Maximal rank" gauge field
produces the s.s.b. of s.i. in a generally covariant theory.

As opposed to the other sections of this paper, the material presented here
has already appeared elswhere. This material is included nevertheless because
it matches perfectly with the other sections and helps to complement and
understand the general phenomenon studied here.

In the first scale invariant model of this kind$^{7}$ , a metric, a dilaton   
field and a "measure field" $\Phi$ are introduced. The unusual modified       
measure$^{8}$ is an object that has the same transformation properties under  
general coordinate transformations as  $\sqrt{-g}d^{4}x$                      
($g = det(g_{\mu \nu})$). $\Phi$ is a density built out of degrees of freedom 
independent of the metric. For example,  given a four index field strength   
(in four dimensions)                                                         
$F_{\mu \nu \alpha \beta} = \partial _{[\mu} A _{\nu \alpha \beta ]}$ as      
$\Phi = \epsilon ^{\mu \nu \alpha \beta} F_{\mu \nu \alpha \beta}$ .

Therefore we  also allow in the          
action, in addition to the                                                    
ordinary measure of integration $\sqrt{-g}d^{4}x$, another one,               
$\Phi d^{4}x$, where $\Phi$ is a density built out of degrees of freedom      
independent of the metric.  One may consider the field strength to be
composed of elementary scalars$^{9}$ . Four in the case of a four index
field strength. For example, given 4-scalars $\varphi_{a}$ (a =                       
1,2,3,4), one can construct the density                                       
\begin{equation}                                                              
\Phi =  \varepsilon^{\mu\nu\alpha\beta}  \varepsilon_{abcd}                   
\partial_{[\mu} \varphi_{a} \partial_{\nu} \varphi_{b} \partial_{\alpha}       
\varphi_{c} \partial_{\beta]} \varphi_{d}                                      
\end{equation}                                                                
     
so that 
\begin{equation}
A _{\nu \alpha \beta} =\varepsilon_{abcd}
 \varphi_{a} \partial_{\nu} \varphi_{b} \partial_{\alpha}
\varphi_{c} \partial_{\beta} \varphi_{d}
\end{equation}

The modified measure idea has been applied to dynamical generation of string
and brane tensions$^{10}$ , to cosmological$^{11}$ and to the fermion family
problem$^{12}$ 
 
   One can allow both geometrical                                        
measures  to enter the theory and consider                                  
\begin{equation}                                                              
S = \int L_{1} \Phi  d^{4} x  +  \int L_{2} \sqrt{-g}d^{4}x                   
\end{equation}                                                                

         Here $L_{1}$ and $L_{2}$ are                                         
$\varphi_{a}$  independent. There is a good reason not to consider            
mixing of  $\Phi$ and                                                         
$\sqrt{-g}$ , like                                                            
for example using                                                             
$\frac{\Phi^{2}}{\sqrt{-g}}$. This is because there  is then an 
invariance (up to the integral of
a total
divergence) under the infinite dimensional symmetry                           
$\varphi_{a} \rightarrow \varphi_{a}  +  f_{a} (L_{1})$                       
where $f_{a} (L_{1})$ is an arbitrary function of $L_{1}$ if $L_{1}$ and      
$L_{2}$ are $\varphi_{a}$                                                     
independent. Such symmetry (up to the integral of a total divergence) is      
absent if mixed terms are present.  A $\frac{\Phi ^{2}}{\sqrt{-g}}$ can
be included in the action, it is consistent with scale invariance, but
would lead to a non zero vacuum energy density in the ground state of the
theory.                                          
                                                                              
        We will study now the dynamics of a scalar field $\phi$ interacting   
with gravity as given by the action (24) with                          
\begin{equation}                                                              
L_{1} = \frac{-1}{\kappa} R(\Gamma, g) + \frac{1}{2} g^{\mu\nu}               
\partial_{\mu} \phi \partial_{\nu} \phi - V(\phi),  L_{2} = U(\phi)           
\end{equation}                                                                
                                                                              
\begin{equation}                                                              
R(\Gamma,g) =  g^{\mu\nu}  R_{\mu\nu} (\Gamma) , R_{\mu\nu}                   
(\Gamma) = R^{\lambda}_{\mu\nu\lambda}, R^{\lambda}_{\mu\nu\sigma} (\Gamma)   
= \Gamma^{\lambda}_                                                           
{\mu\nu,\sigma} - \Gamma^{\lambda}_{\mu\sigma,\nu} +                          
\Gamma^{\lambda}_{\alpha\sigma}  \Gamma^{\alpha}_{\mu\nu} -                   
\Gamma^{\lambda}_{\alpha\nu} \Gamma^{\alpha}_{\mu\sigma}.                     
\end{equation}                                                                
                                                                              
        In the variational principle $\Gamma^{\lambda}_{\mu\nu}$,
$g_{\mu\nu}$, the measure fields scalars                                         
$\varphi_{a}$ and the  scalar field $\phi$ are all to be treated              
as independent variables.                                                     

        If we perform the global scale transformation ($\theta$ =             
constant)                                                                     
\begin{equation}                                                              
g_{\mu\nu}  \rightarrow   e^{\theta}  g_{\mu\nu}                              
\end{equation}                                                                
then the action , is invariant provided  $V(\phi)$     
and $U(\phi)$ are of the                                                      
form                                                                          
                                                                                
\begin{equation}                                                              
V(\phi) = f_{1}  e^{\alpha\phi},  U(\phi) =  f_{2}                            
e^{2\alpha\phi}                                                               
\end{equation}                                                                
and $\varphi_{a}$ is transformed according to                                 
$\varphi_{a}   \rightarrow   \lambda_{a} \varphi_{a}$                         
(no sum on a) which means                                                     
$\Phi \rightarrow \biggl(\prod_{a} {\lambda}_{a}\biggr) \Phi \equiv \lambda   
\Phi $                                                                        
such that                                                                     
$\lambda = e^{\theta}$                                                        
and                                                                           
$\phi \rightarrow \phi - \frac{\theta}{\alpha}$. In this case we call the     
scalar field $\phi$ needed to implement scale invariance "dilaton".           
                                                                           
      Let us consider the equations which are obtained from                 
the variation of the $\varphi_{a}$                                            
fields. We obtain then  $A^{\mu}_{a} \partial_{\mu} L_{1} = 0$                
where  $A^{\mu}_{a} = \varepsilon^{\mu\nu\alpha\beta}                         
\varepsilon_{abcd} \partial_{\nu} \varphi_{b} \partial_{\alpha}               
\varphi_{c} \partial_{\beta} \varphi_{d}$. Since                              
det $(A^{\mu}_{a}) =\frac{4^{-4}}{4!} \Phi^{3} \neq 0$ if $\Phi\neq 0$.       
Therefore if $\Phi\neq 0$ we obtain that $\partial_{\mu} L_{1} = 0$,          
 or that                                                                      
$L_{1}  = M$,                                                                 
where M is constant.             

If the three index field potential  is elementary rather than being 
a composite of  primitive fields $\varphi_{a}$, then still the equation
$L_{1}  = M$, where M is constant is obtained. 
  
 This constant M appears in a self-consistency 
condition of the equations of motion   
that allows us to solve for $ \chi \equiv \frac{\Phi}{\sqrt{-g}}$
(for details see Ref.7.) 
\begin{equation}                                                              
\chi = \frac{2U(\phi)}{M+V(\phi)}.                                            
\end{equation}

        To get the physical content of the theory, it is convenient to go     
to the Einstein conformal frame where                                         
\begin{equation}                                                              
\overline{g}_{\mu\nu} = \chi g_{\mu\nu}                                       
\end{equation}                                                               
 In terms of $\overline{g}_{\mu\nu}$   the non       
Riemannian contribution (defined   as                                         
$\Sigma^{\lambda}_{\mu\nu} =                                                  
\Gamma^{\lambda}_{\mu\nu} -\{^{\lambda}_{\mu\nu}\}$                           
where $\{^{\lambda}_{\mu\nu}\}$   is the Christoffel symbol),                 
disappears from the equations, which can be written then in the Einstein      
form ($R_{\mu\nu} (\overline{g}_{\alpha\beta})$ =  usual Ricci tensor)        
\begin{equation}                                                              
R_{\mu\nu} (\overline{g}_{\alpha\beta}) - \frac{1}{2}                         
\overline{g}_{\mu\nu}                                                         
R(\overline{g}_{\alpha\beta}) = \frac{\kappa}{2} T^{eff}_{\mu\nu}             
(\phi)                                                                        
\end{equation}                                                                
where                                                                         
\begin{equation}                                                              
T^{eff}_{\mu\nu} (\phi) = \phi_{,\mu} \phi_{,\nu} - \frac{1}{2} \overline     
{g}_{\mu\nu} \phi_{,\alpha} \phi_{,\beta} \overline{g}^{\alpha\beta}          
+ \overline{g}_{\mu\nu} V_{eff} (\phi),                                       
V_{eff} (\phi) = \frac{1}{4U(\phi)}  (V+M)^{2}.                               
\end{equation}                                                                
        If $V(\phi) = f_{1} e^{\alpha\phi}$  and  $U(\phi) = f_{2} $  
as required by scale invariance, we obtain from the above equation,

\begin{equation}                                                              
        V_{eff}  = \frac{1}{4f_{2}}  (f_{1}  +  M e^{-\alpha\phi})^{2}        
\end{equation}

 Also a minimum is achieved at zero         
cosmological constant for the case $\frac{f_{1}}{M} < 0 $ at the point         
$\phi_{min}  =  \frac{-1}{\alpha} ln \mid\frac{f_1}{M}\mid $. Finally,        
the second derivative of the potential  $V_{eff}$  at the minimum is          
$V^{\prime\prime}_{eff} = \frac{\alpha^2}{2f_2} \mid{f_1}\mid^{2} > 0$        
if $f_{2} > 0$, so that a realistic scalar field potential, with           
massive excitations when considering the true vacuum state, is achieved in    
a way consistent with the idea of scale invariance.                           
 Since we can always perform the transformation $\phi \rightarrow      
- \phi$ we can                                                                
choose by convention $\alpha > 0$. We then see  that as
$\phi \rightarrow      
\infty, V_{eff} \rightarrow \frac{f_{1}^{2}}{4f_{2}} =$ const.                
providing an infinite flat region.                                                               

The constant $M$ gives, once again s.s.b. of scale invariance.
The conserved currents and the reasons why when $M \neq 0$ these
do not lead to a conserved charge have been studied in Ref. 15.
This explains also the absence of a Goldstone boson in spite of
having s.s.b. of s.i..

\section{Spontaneous Generation of Confining Behavior}
Let us now turn our attention to gauge theories and consider first the
standard pure Yang Mills action
\begin{equation}
S =   \int d^{4}x L,  L = -\frac{1}{4} F^{a}_{\mu \nu}F^{a \mu \nu} 
\end{equation}  

where
\begin{equation}
F^{a}_{\mu \nu} = \partial_{\mu} A^{a}_{\nu} -  \partial_{\nu} A^{a}_{\mu}
+ g f^{abc}A^{b}_{\mu}A^{c}_{\nu}  
\end{equation} 
     
This theory is invariant under the scale symmetry

\begin{equation}
A^{a}_{\mu}(x) \rightarrow A^{a}_{\mu}(x)' = \lambda A^{a}_{\mu}(\lambda x)    
\end{equation} 

Let us rewrite (34) with the help of an auxiliary field $\omega$

\begin{equation}
S =   \int d^{4}x L,  L = \frac{1}{4}\omega^2 
 -\frac{1}{2} \omega \sqrt{ F^{a}_{\mu \nu}F^{a \mu \nu}}  
\end{equation}

upon solving the equation of motion obtained from the variation of
$\omega$ and then replacing into the action (37) we get back (34).

Let us consider now the replacement 
$ \omega \rightarrow$
$ \epsilon ^{\mu \nu \alpha \beta}  \partial _{[\mu} A _{\nu \alpha \beta ]}$  

and consider now the equation of motion obtained from the variation of
$ A _{\nu \alpha \beta }$, which is
\begin{equation}
\epsilon ^{\gamma \delta \alpha \beta} \partial_{\beta} (\omega - 
\sqrt{ F^{a}_{\mu \nu}F^{a \mu \nu}} ) = 0
\end{equation}

which is solved by
\begin{equation}
\omega = \sqrt{ F^{a}_{\mu \nu}F^{a \mu \nu}}  + M
\end{equation}

$M$ being once again a space-time constant which produces s.s.b. of
s.i. and in this case it is furthermore associated with the spontaneous
 generation of confining behavior. Indeed the equations of motion obtained
from (37) in the case $\omega$ is replaced by
$ \epsilon ^{\mu \nu \alpha \beta}  \partial _{[\mu} A _{\nu \alpha \beta ]}$,
  have the form,
 
\begin{equation}
\nabla_{\mu} ( (\sqrt{ F^{b}_{\alpha \beta}F^{b \alpha \beta}} + M)
\frac{F^{a \mu \nu}}{\sqrt{ F^{b}_{\alpha \beta}F^{b \alpha \beta}}}) = 0   
\end{equation}

If we consider the case the argument of the square root is positive, the
 square root itself is defined as positive and $M > 0$ as well, then, 
it is not the case that $\sqrt{ F^{b}_{\alpha \beta}F^{b \alpha \beta}}$
aquires an expectation value. In this case,  
 in the limit of weak field strengths (which dominates the infrared
behavior) eq. (40) gives
\begin{equation}
\nabla_{\mu}(\frac{F^{a \mu \nu}}
{\sqrt{ F^{b}_{\alpha \beta}F^{b \alpha \beta}}}) = 0 
\end{equation} 

Such an equation of motion describe a confining theory$^{16, 17}$. 
For example among other properties, it is known that electric 
monopoles do not exist$^{17}$.
Also string like solutions are a remarkable consequence of such 
equation$^{16, 17}$.

\section{Discussion}
There are several sources for the possible presence of a maximal rank field 
strength. One possibility is of course is that they are provided from the
begining from a microscopic theory. Eleven dimensional supergravity contains
four index field strengths$^{18}$ and although four index field strengths
are not  maximal rank field strengths in eleven dimensions, 
they are  maximal rank field strengths
of the effective four dimensional theory (in the low energy limit).

Another way in which maximal rank field strengths could appear could be a 
totally different mechanism: infrared rather than microscopic.
Let us consider the gauge theory example. There the actions (34) and (37) are
totally equivalent. In (37), the introduction of the $\omega$ field is like
the introduction of intermediate states between two operators, so that
$A^{2} \rightarrow \Sigma  A|n><n|A $ , where $|n>$ is a complete set of 
states.

In the same way, the introduction of the $\omega$ field and the 
$\omega$ field terms in (37) give rise to a gaussian integral in the 
functional integral. The result of doing the functional integral gives us
back the action (34). Thinking of the $\omega$ field as the functional analog
of the states  $|n>$ , we see that with the help of this field, we are 
writing $ F^{a}_{\mu \nu}F^{a \mu \nu}$ 
as the product of  $ \sqrt{F^{a}_{\mu \nu}F^{a \mu \nu}}$ x
 $ \sqrt{F^{a}_{\mu \nu}F^{a \mu \nu}}$ and we insert then the intermediate 
states, represented by $\omega$.

The replacement

\begin{equation}
\omega \rightarrow \epsilon^{\mu \nu \alpha \beta} 
\partial_{[\mu} A_{\nu \alpha \beta]}   
\end{equation}

has then a simple interpretation. As we see from the result of such 
replacement, the connection between $\omega$ and the gauge field 
incorporates now the arbitrary constant $M$ as we can see
from eq. (39). That is the zero energy 
momentum component of  $\omega$ becomes undetermined. Since it is well
known that the nonabelian gauge theory has severe infrared divergences,
which are associated with undetermined  chromomagnetic field strengths
$^{19, 20}$,
declaring this "ignorance" on the zero energy momentum part of the 
"intermediate states" represented by $\omega$  appears to be a well 
motivated step. A proposal for the classification of the  phases of 
a gauge theory, including the
confinement phase, has been discussed in 
terms of the zero energy momentum gauge field content of the vacuum$^{21}$
and  flux tube confinement effects due to the zero energy momentum 
gluons in the vacuum have also been proposed$^{22}$. 
The constant $M$ could have the interpretation then of the
contribution of the zero energy momentum gluons in the vacuum which are
responsible for confinement.

\section*{Acknowledgments}
I would like to thank A.Kaganovich
for discussions.

\end{document}